\documentclass[12pt,preprint]{aastex}
\usepackage{graphics}

\begin{document}
\slugcomment{Submitted to ApJ}

\title{Optical studies of the ultraluminous X-ray source NGC1313 X-2}

\author{Ji-Feng Liu\altaffilmark{1}, Joel Bregman\altaffilmark{2}, Jon
Miller\altaffilmark{2}, and  Philip Kaaret\altaffilmark{3}}

\altaffiltext{1}{Harvard-Smithsonian Center for Astrophysics}
\altaffiltext{2}{University of Michigan}
\altaffiltext{2}{University of Iowa}

\begin{abstract}

NGC1313 X-2 was among the first ultraluminous X-ray sources discovered, and has been a frequent target of X-ray and optical observations.  Using the HST/ACS multi-band observations, this source is identified with a unique counterpart within an error circle of $0\farcs2$.  The counterpart is a blue star on the edge of a young cluster of $\le10^7$ years amid a dominant old stellar population.  Its spectral energy distribution is consistent with that for a Z=0.004 star with 8.5 $M_\odot$ about $5\times10^6$ years old, or for an O7 V star at solar metallicity.  The counterpart exhibited significant variability of $\Delta m = 0.153\pm0.033$ mag between two F555W observations separated by three months, reminiscent of the ellipsoidal variability due to the orbital motion of this ULX binary.

\end{abstract}


\section{INTRODUCTION}

A class of luminous X-ray point sources, usually called ultraluminous X-ray
sources (ULXs), has been established as non-nuclear sources with X-ray
luminosities in the range 10$^{39}$-10$^{41}$ erg/sec. ULXs were discovered in
nearby galaxies first by EINSTEIN (e.g., Fabbiano et al. 1989), ROSAT (e.g.,
Colbert \& Mushotzky 1999), ASCA (e.g., Makishima et al. 2000) and now the
Chandra and XMM X-ray observatories (e.g., Kilgard et al. 2002; Swartz et al.
2004).
ULXs have stirred great excitement in the high energy community because some
may be intermediate mass black holes (IMBHs) of 10$^2$-10$^5$ $M_\odot$ (Miller
\& Colbert 2004). This is because their luminosities exceed the Eddington
limits for 10 $M_\odot$ black holes, and IMBHs are required if the sources are
emitting isotropically at their sub-Eddington levels.
Alternatively, ULXs could be stellar mass black holes with beaming effects,
either mild geometric beaming (King et al. 2001) or relativistic beaming
(Koerding et al. 2001), or with a radiation pressure dominated magnetized
accretion disk where the photon-bubble instability operates and the radiation
can exceed the Eddington limit by a factor of 10-100 (Begelman 2002).

NGC1313 X-2, located about 6$^\prime$ south of the nucleus of NGC1313 at a
distance of 3.7 Mpc (Tully 1988), was among the first ULXs discovered by
EINSTEIN (Fabbiano \& Trinchieri 1987). 
It has been a frequent target of ROSAT, ASCA, Chandra and XMM-Newton
observations, and has shown clear variability of up to a factor of 2 on a
timescale of months (c.f.  Zampieri et al. 2004 and reference therein; Feng \&
Kaaret 2006).
The ULX has a maximum X-ray luminosity of $3\times10^{40}$ erg/sec in 0.3-10
keV, and has switched between high flux state with hard spectrum and low flux
state with soft spectrum, unlike normal Galactic black hole X-ray binaries
(Feng \& Kaaret 2006).
Miller et al. (2003) discovered, along with a power-law component, a cool
accretion disk component of $kT_{in} \approx 160 $eV in its X-ray spectrum,
suggesting an IMBH of $\sim10^3 M_\odot$.
This IMBH interpretation, however, is not unquestionable and the spectrum can
also be fitted by a stellar mass black hole with, e.g., a hot accretion disk
($kT_{in} \sim 2.6 $keV) plus a power-law component (Stobbart et al.  2006).

NGC1313 X-2 has been studied extensively in the optical to understand its
companion and environments.
A deep  $H_\alpha$ image reveals the ULX as located in an elongated
$25^{\prime\prime} \times 17^{\prime\prime}$ supershell, which showed strong [S
II] and [O I] lines with a FWHM of 80 km s$^{-1}$ (Pakull \& Mirioni 2002).
Zampieri et al.  (2004) identified the ULX with a point-like counterpart on an
R image taken with the ESO 3.6m telescope.  Their subsequent efforts
(Mucciarelli et al. 2005) resolved the counterpart into two sources C1 and C2,
with C1 claimed as a B0-O9 main sequence star of $20 M_\odot$ and C2 as a G supergiant
of $\sim 10 M_\odot$. 
Ramsey et al. (2006) studied HST ACS/WFC images and identified the ULX with C1,
which they claimed an early B giant.
Pakull et al. (2006) studied ESO/VLT and SUBARU observations, and found the
counterpart less than 8 $M_\odot$ in an environment of 40-70 Myrs old.
They detected the He II $\lambda4686$ emission line in two blue spectra of the
counterpart taken 3 weeks apart, with a line  shift of 300 km/s.
The He II $\lambda4686$ line indicates the high ionization state for He,
probably due to photoionization by the strong X-ray emission, or photoioniztion
by Wolf-Rayet stars or shocks (e.g., Pakull \& Mirioni 2002).

In this paper we report our studies of NGC1313 X-2 with the HST ACS
observations. In section 2, we describe the observations utilized and the
analysis procedures to identify the counterpart and to compute the photometry.
The ULX is identified with a unique counterpart within an error circle of
$0\farcs2$. Its spectral energy distribution is consistent with that for an O7
V star, and it showed variability above $4\sigma$ between two F555W
observations. The counterpart has an age of $5\times10^6$ years, and is located next to
a loose cluster of blue stars.  We discuss the implications of these results in
Section 3.

\section{DATA ANALYSIS AND RESULTS}

NGC1313 X-2 was observed with HST ACS in four filters F330W, F435W, F555W
and F814W on 2003-11-22, and in filter F555W on 2004-02-22 for program GO-9796.
All observations (Table 1) were downloaded from the HST archive, and calibrated
on the fly with the best calibration files. 

\subsection{Astrometry}

The drizzled ACS/WFC images were used to identify the ULX. These images were
generated by co-adding multiple exposures with the same filter and pointing
with the DRIZZLE package (Fruchter \& Hook 2002), which was able to reject most
cosmic rays plaguing single exposures. The geometric distortion of the camera
was corrected to better than 0.01 pixels (ACS handbook). 
%
The X-ray positions were taken from a 15 ksec Chandra ACIS-I observation (ID
3550), which was downloaded from the Chandra Data Archive and processed with
CIAO 3.3.0. WAVDETECT (Freeman et al. 2002) was run to detect point sources,
and two other X-ray sources (X3 and X4) were detected within $1^\prime$ of the
ULX (Table 2). 


The direct registration of Chandra images on HST images yields a nominal error circle
of $\sim0\farcs8$, as a quadrature sum of the absolute position uncertainties
from Chandra images ($0\farcs6$, Aldcroft et al. 2000) and HST images
($0\farcs5$, HST data handbook).
Luckily, the ULX position on WFC images can be greatly improved with the help
of X4, which was identified with the nucleus of a background galaxy with a size
of $0\farcs5$ as in Figure 1a. 
The correctional shift between the Chandra and HST positions of X4 was applied
to the Chandra position of the ULX to obtain its HST position (Table 2). 
The positional error from this procedure included the ACIS plate scale
variation ($\sim0\farcs03$ for the $36^{\prime\prime}$ separation between X4
and the ULX), the centroiding errors for X-ray sources ($0\farcs17$ for X4 and
$0\farcs01$ for the ULX; Table 2).
These amounted to an error circle of $\le0\farcs2$ around the ULX as shown in Figure 1b.
Within this error circle was only one point-like object, which was the
counterpart candidate C1 in Mucciarelli et al. (2005), and another candidate C2
was unambiguously excluded.
While the highly absorbed X-ray spectrum and highly variable light curve of X4
were consistent with its identification as the nucleus of the background
galaxy, it nevertheless could be any source in the dusty spiral arms. In this
situation, the positional error had contribution from the size of the
background galaxy, and could be as large as $0\farcs5$.

X3 had a very soft spectrum and was identified with a foreground star that was
highly saturated on the WFC images.
The optical position of X3 could be estimated roughly as the crossing of the
two spikes as listed in Table 2.
Despite the uncertainty ($\sim0\farcs5$) of this optical position, the ULX
position obtained through X3 was consistent with that obtained through X4, as
shown on Figure 1b.
Note that the HST coordinates were offset $\sim 2\farcs3$ relative to the
Chandra coordinates. Such an offset was larger than the nominal error of
$0\farcs8$, but was not uncommon, and was real given the successful
identifications of both X4 and X3. In fact, in the study of another ULX in M81,
the HST and Chandra positions for a conspicuous source, SN1993J, were offset by
$\sim2^{\prime\prime}$ (Liu et al. 2002).

\subsection{Photometry}

Photometry for the ULX counterpart and nearby objects was computed on the
drizzled images with the IRAF/DAOPHOT package. 
The drizzled images, originally given in unit of $e^-/s$, were first converted
to $e^-$ per pixel by multiplying the total exposure times (Table 1).
An aperture of 3 pixels was used to extract the flux in the task PHOT, and the
aperture corrections as taken from Sirianni et al. (2005) were applied to
obtain the magnitudes for an ``infinite'' aperture.
A fitting radius of 3 pixels was adopted to fit PSFs
to objects.
We computed the magnitudes both in the VEGAmag and the STmag systems, with the
zeropoints taken from Sirianni et al. (2005).
The photometry results from DAOPHOT were compared to those obtained with the
HSTPHOT package (Dolphin 2000), which were consistent with each other to better
than $\pm$0.1 mag.

%
%
%

Variability of the counterpart was studied with the two F555W observations in
2003 and in 2004, which were separated by three months.
The sources from the two observations were cross identified, which led to
399 bright sources within $10^{\prime\prime}$ of the counterpart.
As shown in Figure 2, most sources did not exhibit significant variability
between the two observations, confirming the stability of our photometry
procedures.
For 318 (i.e., 80\%) of these objects, the variations were less than
2$\sigma_{\Delta m}$, with $\sigma_{\Delta m} \equiv \sqrt{\delta m_1^2 +\delta
m_2^2}$.
Twelve sources exhibited variability above the $4\sigma_{\Delta m}$ significance level,
and the counterpart was one of them with $\Delta m_{F555W} = 0.153 \pm 0.033$,
an amplitude of 15\% at a significance of $4.6\sigma_{\Delta m}$.
%

The multi-band observations in 2003 were used to study the spectral energy
distribution of the ULX counterpart.
The flux densities at the central wavelengths for F330W, F435W, F555W and F814W
were calculated from the STmag in these bands, and plotted in Figure 3 with
corrections for different extinction values based on the extinction law
prescribed by Cardelli, Clayton \& Mathis (1989).
The observations were compared to the absolute magnitudes and colors for blue
stars of O5V, O7V, O9V and B1III spectral types as taken from Schmidt-Kaler
(1982).
If the counterpart is only corrected for the Galactic extinction of 0.11 mag
toward NGC1313 (Schelegel et al. 1998), its F435W and F555W magnitudes are
roughly consistent with those for an O9V star, but its spectral slope is
shallower and its F814W magnitude is much brighter.
If the counterpart is corrected for an extinction of E(B-V) = 0.33 mag, its
overall spectral shape matches perfectly that for an O7V star ($M_V = -5.2$ mag, $\sim30M_\odot$,
$\sim9 R_\odot$).
%
The spectral shape will be over corrected if we apply E(B-V) = 0.44 mag, i.e.,
$n_H = 2.7\times10^{21}$ cm$^{-2}$ as inferred from the power-law fit to the
X-ray spectrum (Miller et al. 2003).
The B1 III star has B and V magnitudes consistent with the counterpart, but its
U magnitude is lower than the observation as shown in Figure 3.
%
Note the properties for the MK spectral types are typical for stars with solar
metallicity in our Solar neighborhood, therefore different from low metallicity
stellar models.

\subsection{Color-Magnitude Diagrams}

The color-magnitude diagrams for the ULX and nearby bright stars were
constructed to estimate their ages. 
Following the advice of Sirianni et al. (2005) against transforming ACS VEGAmag
to standard magnitudes, we constructed the diagrams with VEGAmag and compared
to isochrones computed in the HST ACS/WFC VEGAmag system\footnote{
http://pleiadi.pd.astro.it/isoc\_photsys.02/isoc\_acs\_wfc/index.html} for the
Padova stellar models (Girardi et al. 2002).
Studies of the interstellar medium of NGC1313 indicate a rather low metallicity
of about 0.1-0.2 $Z_\odot$ (Ryder 1993), thus isochrones for Z=0.004 were used
to overlay on the diagrams in Figure 4.
The magnitudes and colors in these diagrams were corrected for the Galactic
extinction of E(B-V)=0.11 mag toward NGC1313.

Two stellar populations were clearly seen from the color-magnitude diagrams, a
young population with ages less than a few  $10^7$ years and an old population with
ages $3-30 \times 10^8$ years.
The ULX counterpart belongs to the young stellar population. 
By comparing its position on the diagrams with those of the stellar models, we
found that the counterpart has an age of $10^7$ years from its F435W and F555W
magnitudes after corrections for E(B-V)=0.11 mag, but has an age of
$3\times10^7$ years from its F555W and F814W magnitudes. 
The two ages have smaller discrepancies for larger extinction corrections, and
converge for E(B-V) = 0.33 mag, for which the counterpart has an age of
$5\times10^6$ years, an initial/current mass of $\sim52/8.5 M_\odot$, and a
radius of $\sim7R_\odot$. 
%
%
If this extinction is applied to all stars, the young population has ages of
$5-10 \times 10^6$ years, and the old population has ages of $2-20 \times 10^8$
years.

The two stellar populations are obvious on the accompanying $50^{\prime\prime}
\times 50^{\prime\prime}$ true-color image centered at the ULX counterpart
(Figure 7).
The image was constructed from the F435W, F555W, and F814W observations in
2003, and the bright cosmic rays not removed by the DRIZZLE package were
manually removed.
The field was dominated by old red faint stars, with scattered young blue
stars.  A few tens of bright blue stars were loosely clustered within an
irregular region of $10^{\prime\prime} \times 5^{\prime\prime}$ (i.e.,
$180pc\times 90pc$) as outlined on the image. 
The counterpart is the marked bright blue star located to the east-south edge
of this loose cluster.
The cluster region is much larger than for typical Galactic open clusters
($\le10pc$), and the member stars may be the products of one star formation
episode that are not bounded gravitationally.
We note that this loose cluster consists of only blue stars, and no red
supergiants.
For the adopted stellar models with $Z=0.2Z_\odot$, only stars less massive
than $20 M_\odot$ can evolve to red supergiants in $\sim10^7$ years or longer.
The absence of red supergiants indicates the open cluster is younger than
$\sim10^7$ years, regardless of the extinction correction. This is consistent
with the results from comparison with isochrones.

\section{DISCUSSION}


Different efforts to identify the counterpart for NGC1313 X-2 have now
converged upon the same blue star.
By registering X-2 onto an ESO 3.6m R image with respect to SN1978K at
$\sim5^\prime$ away, Zampieri et al. (2004) identified a point-like object
within the $1\farcs4$ error circle, which was later resolved into two stars, C1
and C2, on ESO VLT images (Mucciarelli et al. 2005).
Ramsey et al. (2006) used stars in the USNO B1.0 catalog to determine
astrometric solutions for the HST ACS/WFC images to an rms accuracy of $0\farcs5$,
and identified C1 as the counterpart.
In this paper, we registered X-2 onto the HST ACS/WFC images with respect to
two nearby X-ray sources that were both identified optically, and confirmed C1
as the counterpart within an error circle of $0\farcs2$.
The chance for this blue star to be a bright interloper is negligible
($<10^{-2}$) given the small size of the error circle and the local stellar
surface density.
Furthermore, the blue star exhibited significant variability between two F555W
observations, as expected if this blue star is the secondary of the X-ray
binary.
We conclude that the blue star C1 is the true counterpart for NGC1313 X-2.


Ramsey et al. (2006) studied the counterpart with the transformed B and V
magnitudes and claimed it a 7-9 $M_\odot$ early B giant. 
They studied the nearby stars without any extinction correction and concluded
that the stellar population was at least $10^7$ years old without any O stars.
In this work, two more bands, F330W and F814W, were included in the spectral
energy distribution of the counterpart, and it was found that an early B giant
such as B1III does not match the overall spectral shape of the counterpart.
The counterpart is brighter and hotter than a B1III giant, and should have a
spectral form of $\nu^{-2}$ in the observed optical bands, i.e., the Jeans tail
for a blackbody spectrum.
To correct the observed spectral shape to $\nu^{-2}$, an extinction value of
E(B-V) = 0.33 mag is required.
This value is, reasonably, lower than the value derived from X-ray absorption
(0.44 mag); it is higher than the Galactic value (0.11 mag), suggestive of
additional extinction in NGC1313 and the very vicinity of NGC1313 X-2.
With this extinction correction, the counterpart has magnitudes and colors for
a $Z=0.2Z_\odot$ star with an age of $5\times10^6$ years, a current mass of
$\sim8.5 M_\odot$, and a radius of $\sim7R_\odot$.
Incidentally, the corrected magnitudes and colors are also consistent with those
for an O7V star with $Z=Z_\odot$, a mass of 30 $M_\odot$, and a radius of 9
$R_\odot$.
This identification of the ULX as an O7V star joins previous ULX
identifications as massive OB stars (e.g., Liu et al. 2002; Liu et al. 2004;
Kaaret et al. 2004) to corroborate the ULX scenario in which the massive
secondary is overflowing its Roche Lobe to supply the high accretion rate
required for the high X-ray luminosity (king et al. 2001).


Pakull et al. (2006) speculated that the optical light from the X-ray
illuminated the accretion disk dominated that from the secondary which they
claimed a B giant.
They argued that the X-ray illuminated accretion disk can reach $M_V \sim -5$
mag for a stellar mass black hole with $L_X/L_{E} \sim 100$ and $P\sim200$
hours based on the van Paradijs \& McClintock $\Sigma - M_V$ diagram (1994)
derived for Galactic low mass X-ray binaries, if the relation still holds for
$L_X>10^{40}$ erg/s.
For an X-ray illuminated accretion disk, the temperature depends on the radius
$R$ roughly as $R^{-0.5}$ (e.g., Frank, King \& Raine 2002; Raymond 1993), and
decreases to about 4000 K at $R \le 9R_\odot$. The resulted spectral energy
distribution takes a form of $\nu^{-1}$ in the observed bands, redder than the
O star spectrum ($\nu^{-2}$) by a color difference $\Delta(B-V) \sim0.25$
mag\footnote{For the O star spectrum, $B-V = 2.5 lg(f_V/f_B) - 2.5
lg(f_{V0}/f_{B0}) = 2.5 lg(\lambda_V^2/\lambda_B^2) - 2.5 lg(f_{V0}/f_{B0})$;
for the accretion disk spectrum, $B-V = 2.5 lg(f_V/f_B) -2.5 lg(f_{V0}/f_{B0})
= 2.5 lg(\lambda_V/\lambda_B) -2.5 lg(f_{V0}/f_{B0})$. Thus $\Delta(B-V) = 2.5
lg(\lambda_V/\lambda_B)
\approx 0.25$ mag for $\lambda_V = 0.55$ nm and  $\lambda_B = 0.435$ nm.}.
The presence of significant accretion disk light would modify the
classification of the secondary and its estimated properties.
In Section 2, we corrected the observed spectral energy distribution to that
for an O7V star ($M_V=-5.2$ mag) by an extinction of E(B-V) = 0.33 mag, an
extra E(B-V)=0.22 mag in addition to the Galactic E(B-V)=0.11 mag.
Alternatively, the observed spectral energy may be interpreted as 12\% from the
secondary ($\nu^{-2}$) plus 88\% from the accretion disk ($\nu^{-1}$) corrected
only by the Galactic extinction E(B-V)=0.11 mag, or 56\% from the secondary
plus 44\% from the accretion disk corrected by an extinction of E(B-V)=0.22
mag.
In the first alternative, the secondary would have $M_V = -2.9$ mag, similar to
a B1V star ($M_V=-3.2$ mag) or slightly later. In the latter, the secondary
would have $M_V=-4.6$  mag, similar to a B1III giant ($M_V=-4.4$ mag).

We suggest the accretion disk light does not dominate the secondary light.
The first of the above alternatives is not reasonable because there must be
some extinction intrinsic to NGC1313. 
The second alternative is not true either, or else the composite spectrum would
show a drop in F330W due to the cool B1III giant, which we did not observe.
In fact, the same $\Sigma - M_V$ relation predicts a $M_V\sim-2.2$ mag for the
accretion disk around an IMBH of $\sim10^2 M_\odot$ and a period of $\sim50$
hours as shown later, $<$ 10\% of the total observed light.
Copperwheat et al (2006) carried out detailed simulations for seven ULX
systems, and found that the light from the disk, depending on factors such as
the disk size and the inclination angle, is generally much less than that from
the secondary.
Similarly, we expect that the optical light from the accretion disk is
insignificant as compared to the massive secondary star.
Nonetheless, the relative fraction of the optical light from the accretion disk
can be measured with multiple simultaneous optical and X-ray observations. 
Assume the total optical light $L_{opt} = L_{secondary} + L_{disk}$, with
$L_{secondary}$ constant in all observations, and $L_{disk} = C\times L_X$ in
linear proportion to the X-ray luminosity $L_X$.
In principle we can solve or best fit for $L_{secondary}$ and the linear factor
$C$ if we have multiple (2+) observations in which the X-ray luminosities vary
significantly.
%

%

\subsection{The association with a young cluster}


We emphasize that there is a young stellar population of $\le10^7$ years in
additional to the apparent old population around NGC1313 X-2.
Red stars from the old population are dominant in this sky region, while blue
stars are rarer.
Careful inspection of the accompanying true-color image shows that these blue
stars are loosely clustered in a $\sim10^{\prime\prime} \times
5^{\prime\prime}$ region that is not gravitationally bounded. 
The member stars may have been generated from the same star formation episode
that is still active.
%
%
Given the low surface density of this young cluster, it is easily missed,
leading to claims that the whole stellar population is old.
This is particularly true for short exposures and low resolution images, and
may lead to an underestimate of the ULXs that are identified with young massive
stars.
%


There are implications on the formation of NGC1313 X-2 given its location
nearby, but not in the center of, a young cluster that is not gravitationally
bounded.
If NGC1313 X-2 has radiated below its Eddington luminosity, its peak luminosity
of  $3\times10^{40}$ erg/sec will suggest a black hole of $\ge200M_\odot$.  
Such an IMBH cannot be formed from massive stars with large mass loss due to
stellar winds incurred by metal lines, but can be formed through mergers in
dense clusters.
One scenario is the merging of stellar mass black holes in the dense core of
globular clusters (Miller et al. 2002). The chance for thus formed IMBHs below
300 $M_\odot$ to be kicked out of the cluster centers is not negligible, but
the time scale is much longer than $10^8$ years to form a 200 $M_\odot$ IMBH
(Gultekin et al. 2004).
Another scenario is the merging of massive stars in super star cluster of
$\sim10^6 M_\odot$ (e.g., Portegies Zwart \& McMillan 2002). Frequent stellar
captures and mergers lead to super stars of $\sim10^3 M_\odot$ that
subsequently form IMBHs via direct collapses or supernova explosions in a few
million years. The resulted IMBHs, however, tend to sit in the centers of super
clusters, inconsistent with X-2's location outside a young cluster.

Another formation scenario for NGC1313 X-2 is the merging of proto stars in
proto clusters as suggested by Soria (2005). 
Stellar captures and mergers are enhanced by proto stellar disks/envelops and
low stellar velocities, and super stars of $\sim10^3 M_\odot$ can form in less
than a million years in small proto clusters of $\sim 10^4 - 10^5 M_\odot$.
Subsequent explosions of these super stars will unbound the proto clusters, and
form IMBHs of a few hundred $M_\odot$ 
in weakly bounded or unbounded clusters, usually off-center as found for
NGC1313 X-2. 
This formation mechanism works best in low Z environments 
and in colliding or tidally interacting systems 
as exemplified by the Antennae (Fabbiano et al. 2002), Cartwheel (Gao et al.
2003), NGC7714/7715 (Smith et al. 2004) and M99 (Soria et al. 2005).
NGC1313 X-2 is, incidentally, in an environment of low metallicity ($\sim
0.1-0.2 Z_\odot$) in the southern satellite regions of NGC1313, which could be
a tidally disrupted companion galaxy (Sandage \& Brucato 1979) or the result of
collision of huge HI clouds with the disk of NGC1313 (Marcelin \& Gondoin
1983).

\subsection{Optical variations}


The counterpart exhibited variability of $\Delta m = 0.153 \pm 0.047$ mag
(3.3$\sigma$) between two ACS/WFC F555W observations separated by three months.
Most probably, this variability reflects the orbital motion of an inclined
binary system for NGC1313 X-2.
Indeed, the variability is about the level of ellipsoidal variability expected
for a high-mass secondary that approximately fills its Roche lobe.
For example, the amplitude is $\approx 0.17$ mag for an O5V secondary, a 10
$M_\odot$ black hole, $L_X=10^{40}$ erg/s, and $cosi=0.5$ (Copperwheat et al.
2005).

It is unlikely that this variability is caused by the intrinsic variability of
a pulsating star as the secondary.
The observed variability and the blue color may be possible for a $\beta$
Cephei star that varies by 0.01 - 0.3 mag with periods of 0.1 - 0.6 days, or a
Wolf-Rayet star or a Luminous Blue Variable (LBV) which exhibit day-to-day
microvariations of 0.1-0.2 mag and ``normal'' variations of 1-2 mag on
timescales longer than months.
However, the secondary cannot be a Wolf-Rayet star because no prominent
emission lines for Wolf-Rayet star were shown in VLT observations (Pakull et
al. 2006). The secondary cannot be a $\beta$ Cephei star or a LBV because the
observed luminosity ($M_V\sim-5$) is high for a typical $\beta$ Cephei star
($M_V\sim-2$), but low for a typical LBV ($M_V\sim-7$).

On the other hand, the optical variability might have sizable contribution from
the accretion disk in presence of substantial X-ray variability (e.g., flares).
Fortunately, the change of the optical spectra will enable us to quantify this
contribution, because the spectrum of an O type secondary is bluer than the
accretion disk by a color difference of $\Delta (B-V) = 0.25$ mag.
For example, if the accretion disk light increases from 5\% to 20\% of the
secondary light, the spectrum will get redder by $\Delta(B-V) \approx$ 0.04
mag.
Such a color difference can be detected by HST observations with 1\%
photometric accuracy.

A well designed photometric monitoring program can sample the ellipsoidal light
curve and  obtain the orbital period.
%
The period of NGC1313 X-2 can be calculated as a function of the black hole
mass ($M_1$) for the known secondary mass ($M_2$) and size ($R_2$).
The Roche lobe size $R_{cr}$ of the secondary is solely determined by the
orbital separation $a$ and the mass ratio $q = M_2/M_1$ as $R_{cr} = a f(q)$.
If the secondary is filling its Roche lobe, i.e., $R_2 = R_{cr}$, we can
compute the orbital separation as $a = R_2/f(q)$, and subsequently calculate
the period as $P^2 = 2\pi {a^3 \over G(M_1+M_2)}$.
The predicted periods for different primary masses are shown in Figure 5, in
which $f(q)$ values as tabulated in Kopal (1959) are used.
The analytic approximations to these tabulated values by Paczynski (1971) and
by Eggleton (1983) were also tried, which gave periods different from Kopal of
less than 5\% as shown in Figure 5.


The predicted periods are largely determined by the secondary properties.
Adopting Paczynski's approximation for $q<0.8$, $f(q) = 0.46224 ({q \over 1 +
q})^{1/3}$, the period can be expressed as $P = \sqrt{110/\rho} $ hr
independent of the primary mass, in which $\rho$ is the mean density of the
secondary in unit of g cm$^{-3}$ (Frank, King \& Raine 2002).
For massive black holes ($q<0.8$), the period is $\sim44$ hours for a
secondary of an O7V star ($30M_\odot,9R_\odot$), $\sim115$ hours for a B0III
giant ($20M_\odot,15R_\odot$), and $\sim290$ hours for a B0I supergiant
($25M_\odot,30R_\odot$).
Thus, the observed period can be used to test possible classifications of the
secondary.
On the other hand, if the secondary mass and size can be determined by other
means, the observed period can be used to constrain the black hole mass.
In the case of NGC1313 X-2, the secondary mass and size are determined to be
8.5 $M_\odot$ and 7 $R_\odot$ by comparing to $Z=0.02Z_\odot$ stellar models.
We expect a period of $\sim56$ hours for massive black holes, and shorter
periods for primaries less massive than $\sim15 M_\odot$, i.e., a stellar mass
black hole for this ULX (Figure 5).


The orbital velocities around the mass center for the Roche lobe filling system
NGC1313 X-2 are shown in Figure 6 for different primary masses. 
For the secondary of 8.5 $M_\odot$ and 7 $R_\odot$, the velocity increases from
a few tens to a few thousands km/sec for black hole masses from a few to a few
thousand $M_\odot$. 
For an IMBH of a few hundred $M_\odot$, the velocity change is about $10^3$
km/sec, easily detectable with low spectral resolutions of a few \AA \  at
optical wavelengths.
The orbital velocity for the black hole (i.e., the accretion disk) decreases from
$\sim300$ to a few km/sec for increasing black hole masses.
Note the radial velocity is smaller than the orbital velocity by $sini$ for an
inclination angle $i$.

Pakull et al. (2006) detected broad He II $\lambda4686$ emission lines with
FWHMs of $\sim600$ km/s in two ESO VLT spectra  of NGC1313 X-2, which they
attributed to the X-ray illuminated accretion disk.
The He II line appeared to have moved by 300 km/sec, and may indicate the
system has a stellar mass black hole since the orbital velocity change is very
small for very massive black holes (Figure 6).
However, the He II line from an X-ray illuminated accretion disk is usually
broader, e.g., about $\sim1200$ km/s for an inclination angle of $\sim30^\circ$
(Raymond 1993).
The emission line will be narrower for smaller inclination angles (i.e,
face-on), but the full amplitude radial velocity change will be reduced to much
lower than 300 km/s.
We suggest that the He II line comes from the X-ray photoionized
nebula/secondary instead of the accretion disk. The line shift may be caused by
noise given the low spectral resolution and low signal-to-noise ratios of the
observations.
Further observations that sample different orbital phases, with higher spectral
resolution and higher signal-to-noise ratio,  will help to interpret the He II
line and place tight constraints on the black hole mass.

\acknowledgements

We would like to thank Jeffery McClintock, Roberto Soria, Kayhan Gultekin,
Rosanne Di Stefano, Manfred Pakull, Andrew King, Miriam Krauss, Luca Zampieri,
Michael Garcia for helpful discussions, and thank Leo Girardi and  Marco
Sirianni for their help on isochrones. JFL acknowledges the support for this
work provided by NASA through the Chandra Fellowship Program, grant PF6-70043.



\begin{deluxetable}{lllllllc}
\tablecaption{The HST ACS observations for NGC1313 X-2}
\tablehead{
\colhead{ID} & \colhead{Filter} & \colhead{ExpT} & \colhead{DATE} & \colhead{ACor} & \colhead{ Z$_{VEGA}$} & \colhead{Z$_{ST}$} &  VEGAmag \\
}

\startdata
j8ola2010  & HRC/F330W & 2760 & 2003-11-22 & 0.420 & 22.904 & 23.026 & 22.037$\pm$0.021 \\
j8ol02040  & WFC/F435W & 2520 & 2003-11-22 & 0.277 & 25.779 & 25.157 & 23.470$\pm$0.017 \\
j8ol02030  & WFC/F555W & 1160 & 2003-11-22 & 0.249 & 25.724 & 25.672 & 23.625$\pm$0.026 \\
j8ol02010  & WFC/F814W & 1160 & 2003-11-22 & 0.292 & 25.501 & 26.776 & 23.640$\pm$0.043 \\
j8ol06010  & WFC/F555W & 2240 & 2004-02-22 & 0.249 & 25.724 & 25.672 & 23.472$\pm$0.021 \\

\enddata

\tablecomments{The columns are (1) exposure ID, (2) filter, (3) total exposure
in seconds, (4) observation date, (5) aperture correction in magnitude, (6)
zeropoint for VEGAmag, (7) zeropoint for STmag, and (8) VEGAmag for the
counterpart. }

\end{deluxetable}


\begin{deluxetable}{llll|ll}
\tablecaption{The optical counterpart for NGC 1313 X-2}
\tablehead{
\colhead{} & \multicolumn{3}{c}{ACIS-I} \vline& \multicolumn{2}{c}{ACS/WFC} \\
\colhead{Object} & \colhead{R.A.} & \colhead{Decl.} & \colhead{error} \vline& \colhead{R.A.} & \colhead{Decl.} 
}

\startdata
X4            &03:18:25.250&-66:36:36.04 & $0\farcs17$ & 03:18:24.911 &-66:36:37.09 \\
ULX$^a$       &03:18:22.238&-66:36:03.49 & $0\farcs01$ & 03:18:21.900 &-66:36:04.54 \\
\hline
X3            &03:18:29.287&-66:35:36.53 & $0\farcs20$ & 03:18:29.015 &-66:35:37.57 \\
ULX$^b$       &            &             &             & 03:18:21.967 &-66:36:04.54 \\

\enddata

\tablecomments{(a) This ULX position on ACS/WFC images is derived relative to
the X4 position on ACS/WFC images. (b) This ULX position on ACS/WFC images is
derived relative to the X3 position on ACS/WFC images. }

\end{deluxetable}

\clearpage

\begin{figure}
\plottwo{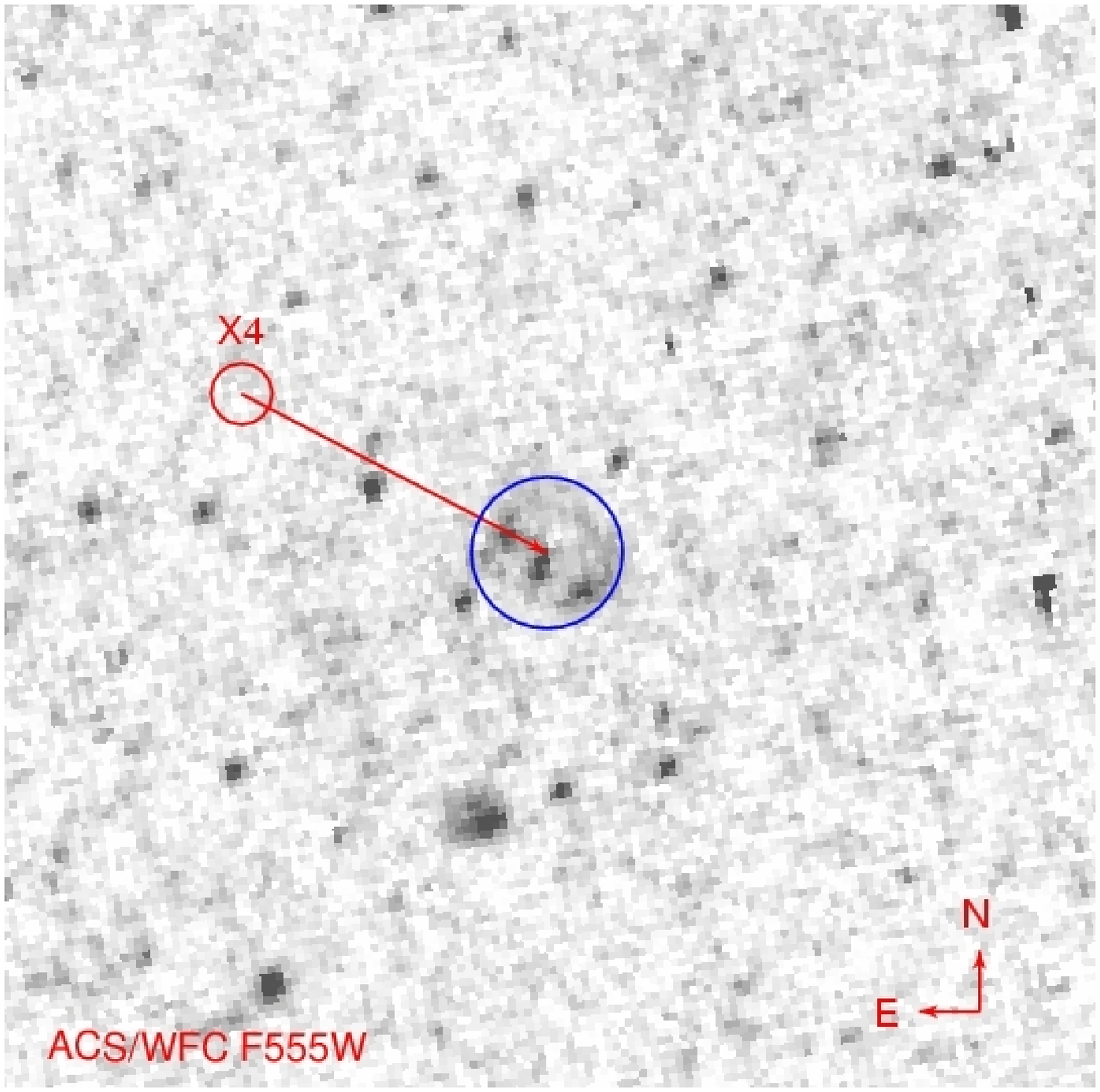}{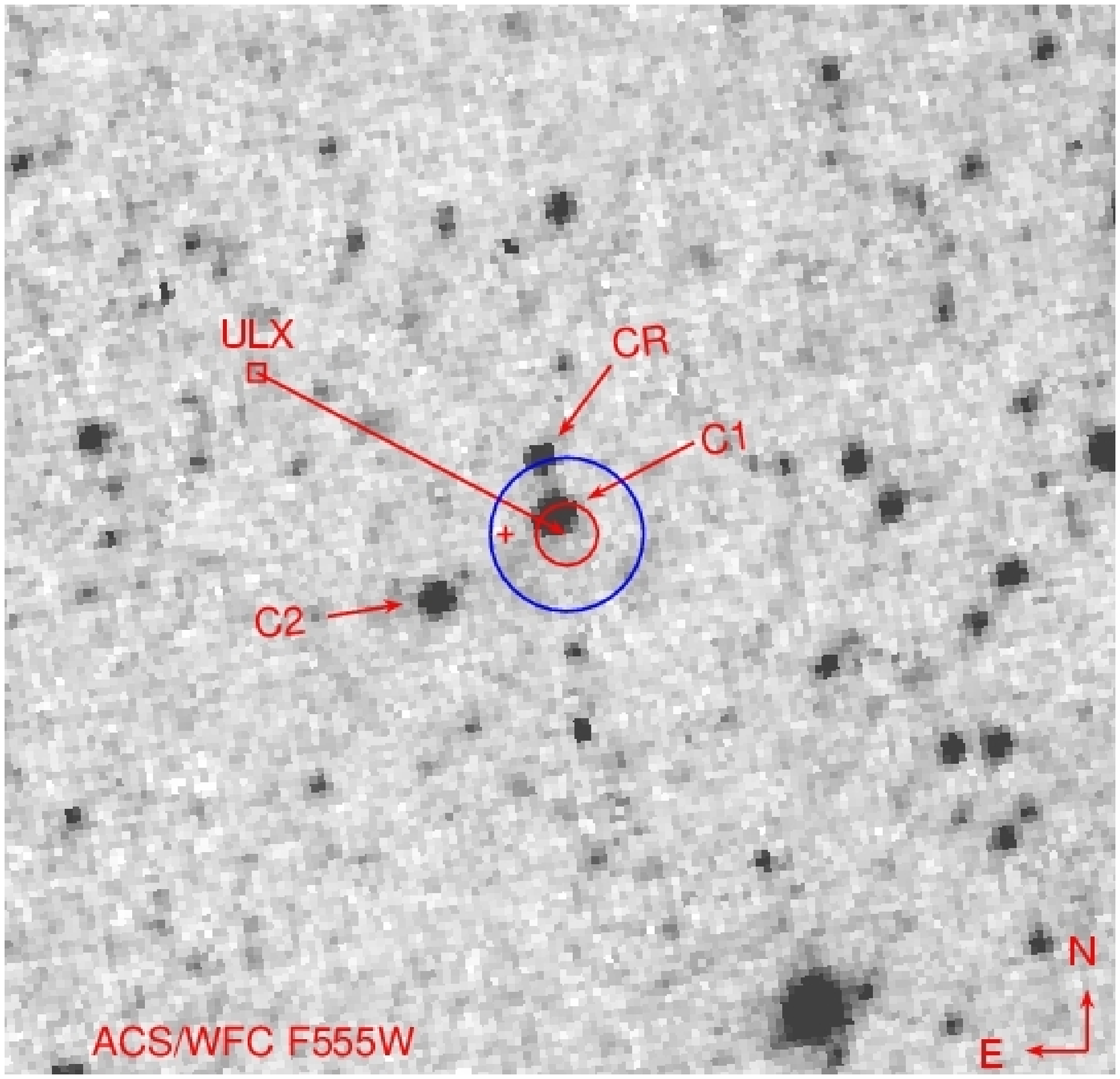}
\caption{The identification of the ULX and X4 on the WFC F555W image. (a) X4
was identified with the nucleus of a background galaxy as enclosed by a
$0\farcs5$ circle. The smaller $0\farcs17$ circle denotes the Chandra position
of X4.  The line segment indicates the relative offset of $2\farcs3$ between
Chandra and HST positions.  (b) The ULX was identified with a unique object
labeled as 'C1' through the identification of X4.  Around 'C1", the small error
circle had a radius of $0\farcs2$, and the large error circle $0\farcs5$.  The
line segment indicating the offset between Chandra and HST positions was based
on the identification of X4.  The cross on the inner edge of the $0\farcs5$
circle indicates the ULX position obtained through the identification of X3 as
a saturated foreground star. The object to its north was a cosmic ray only
appeared in the 2003 F555W observation. Both panels show
$7^{\prime\prime}\times7^{\prime\prime}$ regions.  }

\end{figure}

\begin{figure}
\plotone{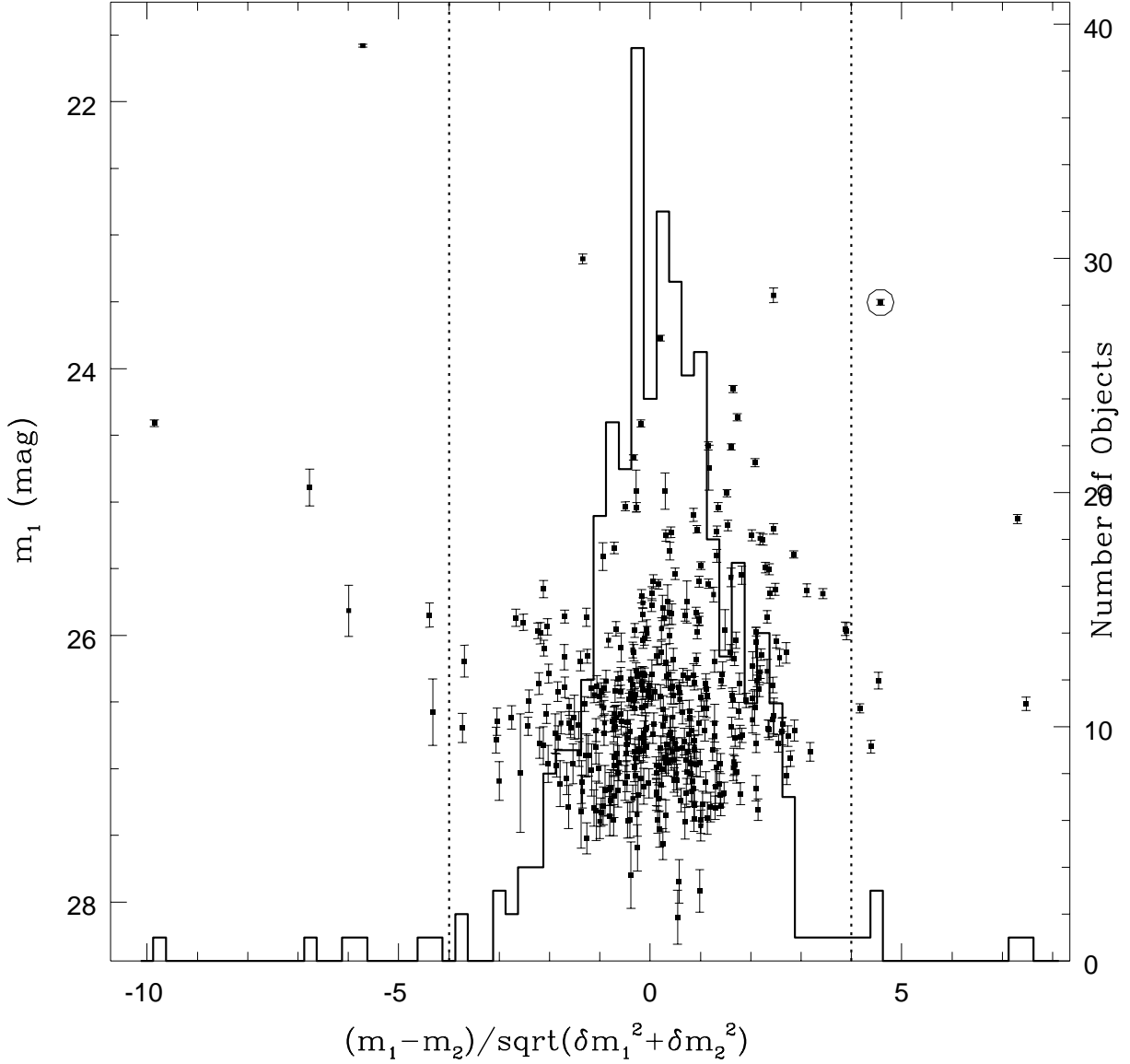}
\caption{Variability between two WFC F555W observations for objects within
10$^{\prime\prime}$ of NGC1313 X-2. The error bar size is equal to $\sigma_{\Delta m}
= \sqrt{\delta m_1^2+\delta m_2^2}$.  The error bars are generally smaller for
brighter objects; occasionally bright objects have large error bars because
they are in unresolved stellar fields.  The counterpart for NGC1313 X-2 is
highlighted by a large circle. The histogram is for the number of objects per
0.25 in $(m_1-m_2)/\sigma_{\Delta m}$.  The dotted lines denote $m_1 - m_2 =
\pm4\sigma_{\Delta m}$. }

\end{figure}

\begin{figure}
\plotone{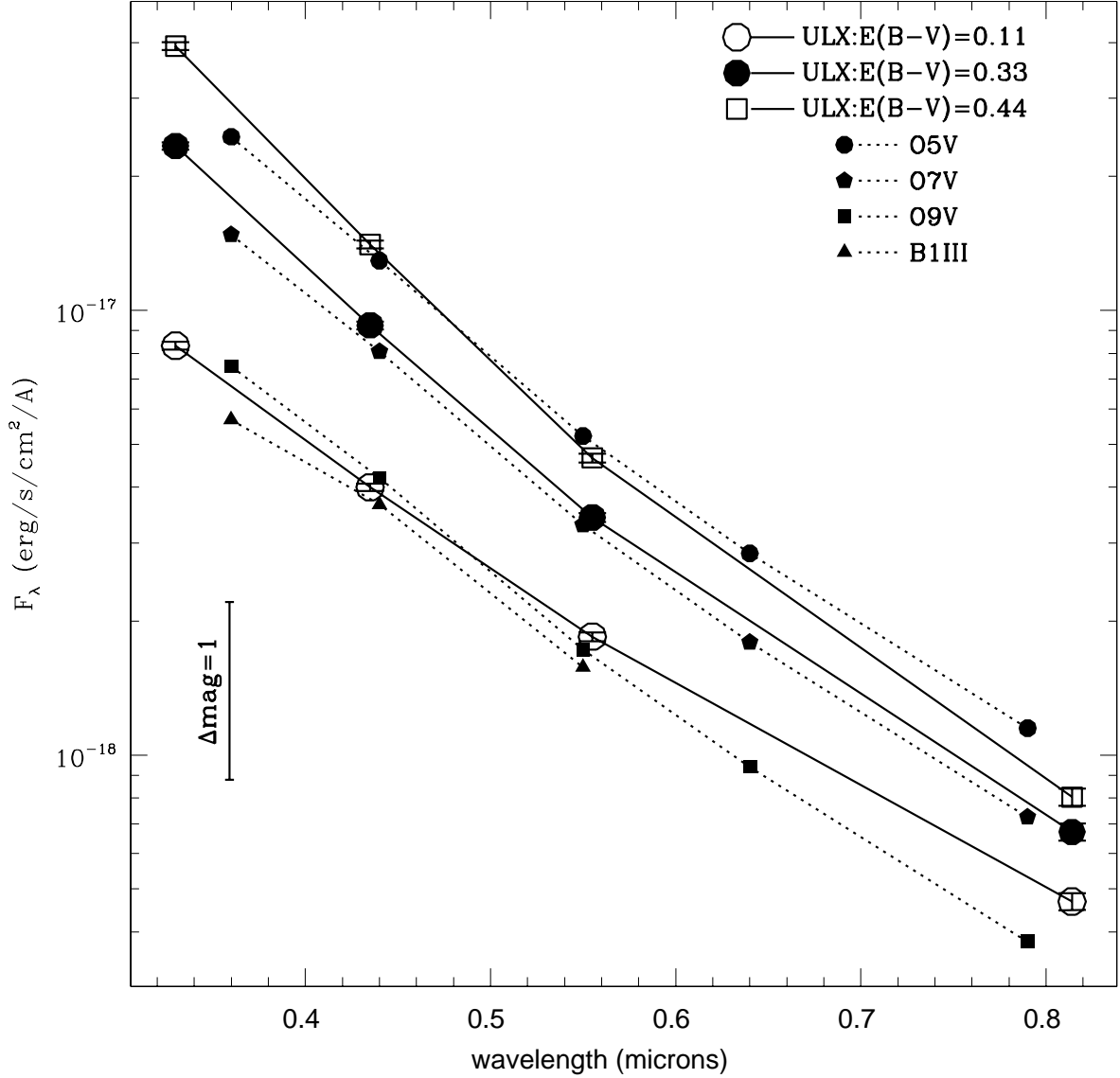}
\caption{The spectral energy distributions for the ULX counterpart in
comparison with standard stars of different MK spectral types. The absolute
magnitudes of these spectral types are taken from Schmidt-Kaler (1982). All
magnitudes are converted to fluxes to enable comparison between HST and UBVRI
systems. The magnitudes/fluxes in filters F330W, F435W, F555W, and F814W were
corrected for E(B-V)=0.11 mag ($A = 0.561,0.455,0.335,0.196$, respectively),
E(B-V)=0.33 mag ($A = 1.684,1.366,1.005,0.587$, respectively), and E(B-V)=0.44
($A = 2.245,1.821,1.340,0.783$, respectively) assuming a Galactic extinction
law.  The overall spectral shape of the counterpart dereddened with E(B-V)=0.33
mag best matches the shape of an O7V star.}

\end{figure}

\begin{figure}
\plottwo{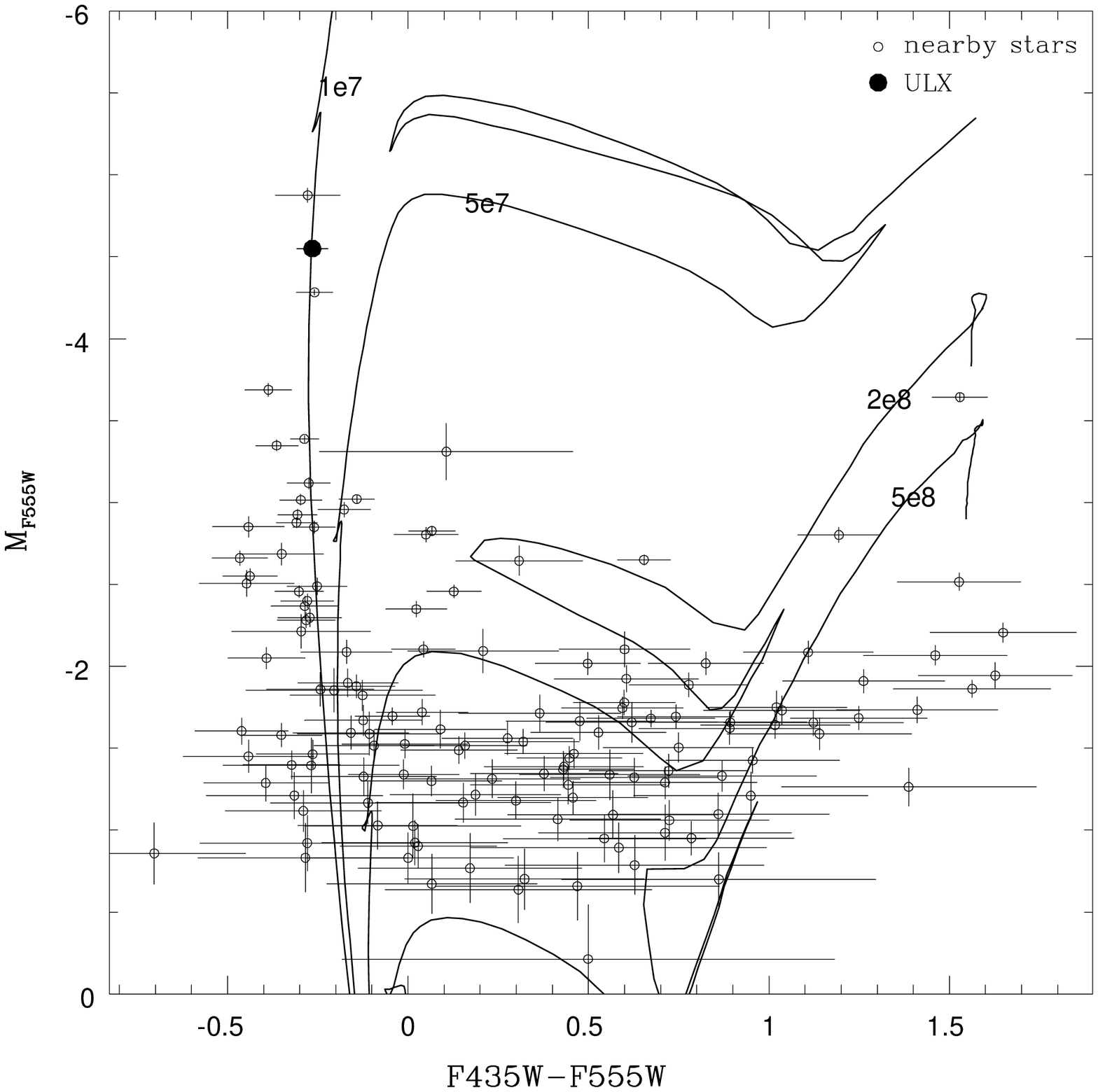}{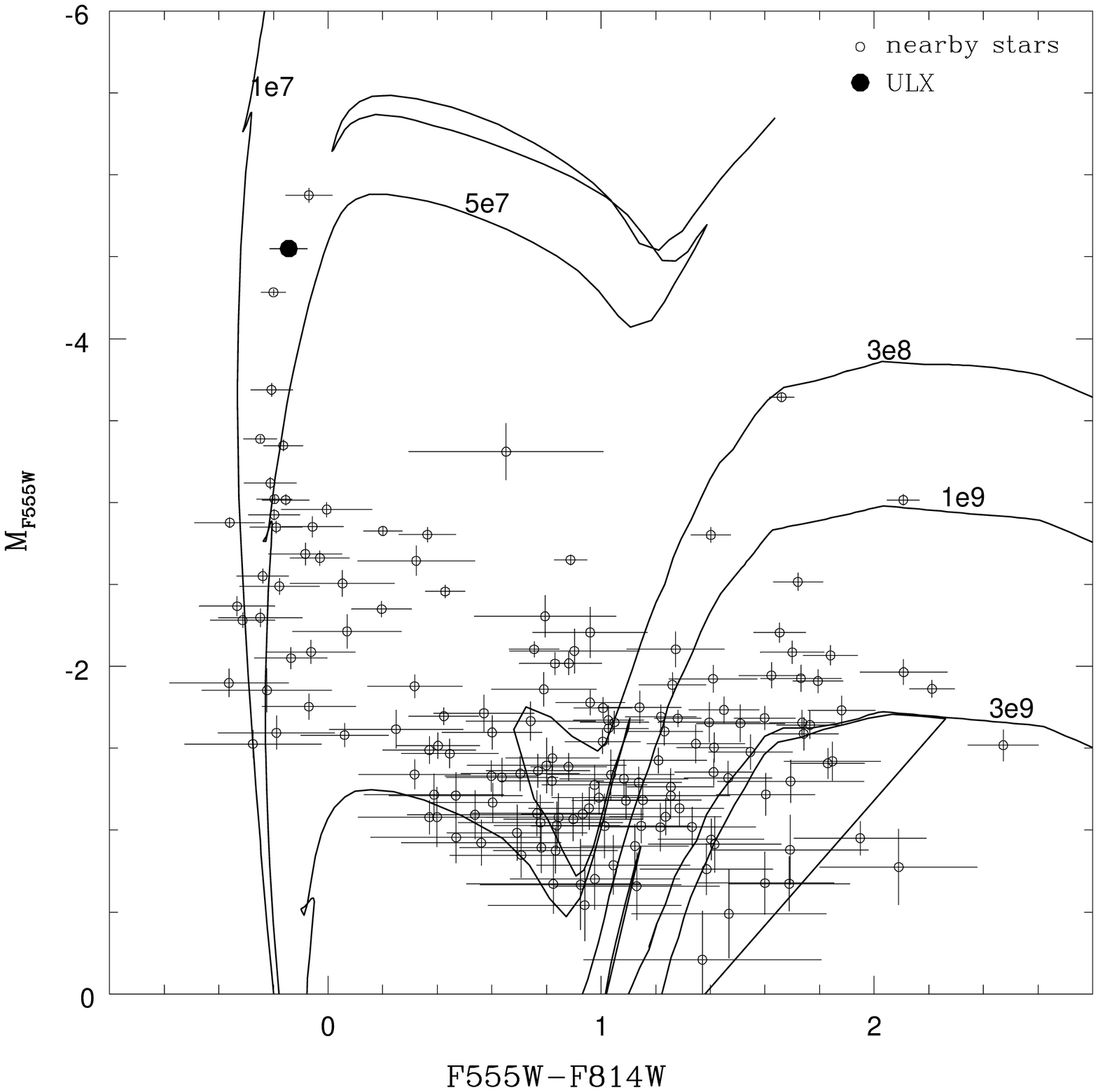}
\caption{The color-magnitude diagrams for the ULX counterpart and nearby bright
stars. Both the data and isochrones are in the HST ACS/WFC VEGAmag system. The
magnitudes are corrected for the Galactic extinction of E(B-V)=0.11 mag. (a)
$M_{F555W}$ vs. F435W-F555W. The ages for the isochrones are $10^7$,
$5\times10^7$, $2\times10^8$, and $5\times10^8$ years, respectively. (b)
$M_{F555W}$ vs. F555W-F814W. The ages for the isochrones are $10^7$,
$5\times10^7$, $3\times10^8$, $10^9$, and $3\times10^9$ years, respectively. }

\end{figure}

\begin{figure}
\plotone{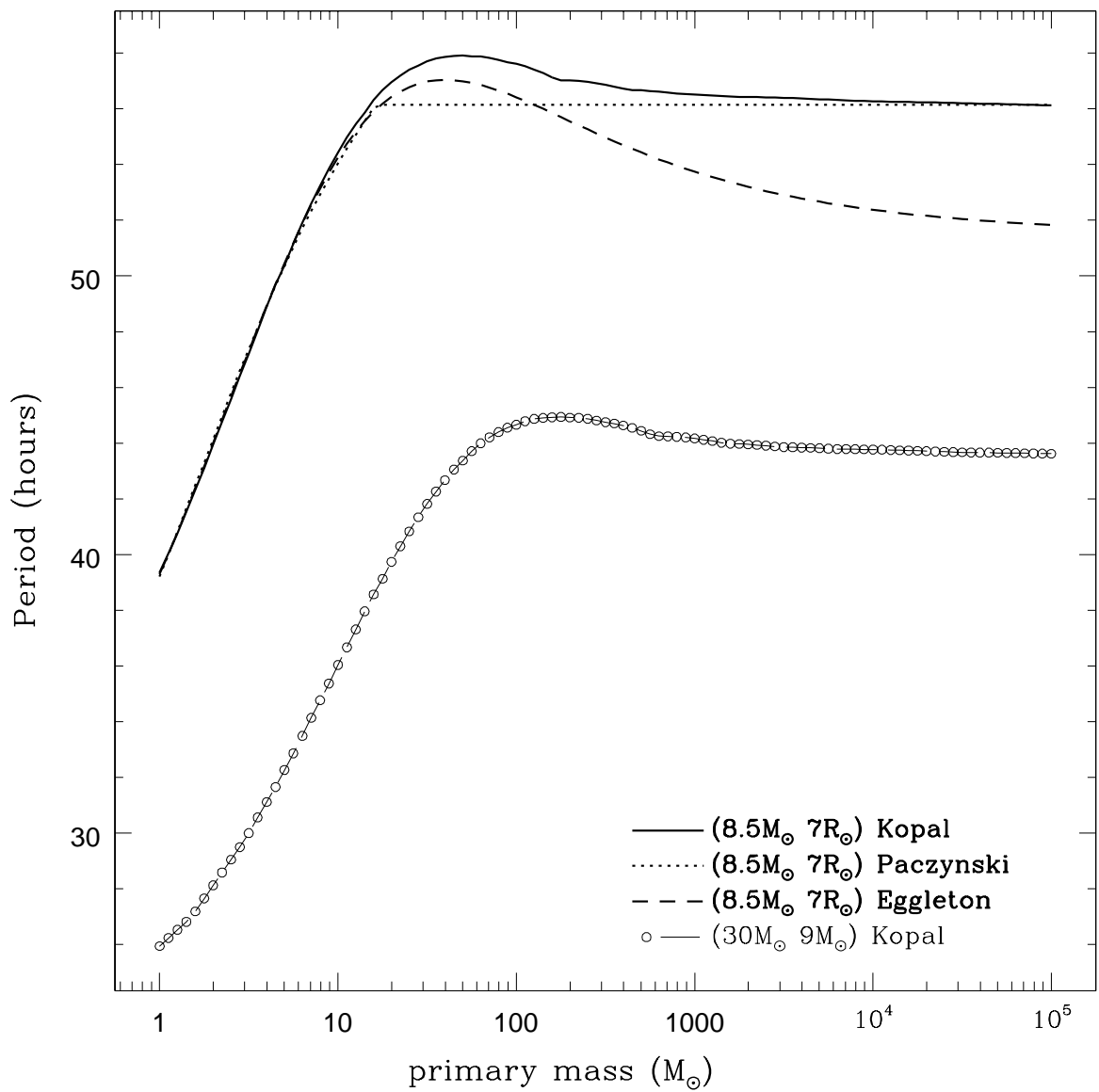}
\caption{The predicted periods for different primary masses for a
$Z=0.2Z_\odot$ secondary ($8.5M_\odot,7R_\odot$) and for an O7V secondary
($30M_\odot,9R_\odot$) with solar metallicity.  The periods are calculated
using the Roche lobe sizes $f(q)$ as tabulated by Kopal, and the analytic
approximations by Paczynski and Eggleton.  }

\end{figure}

\begin{figure}
\plotone{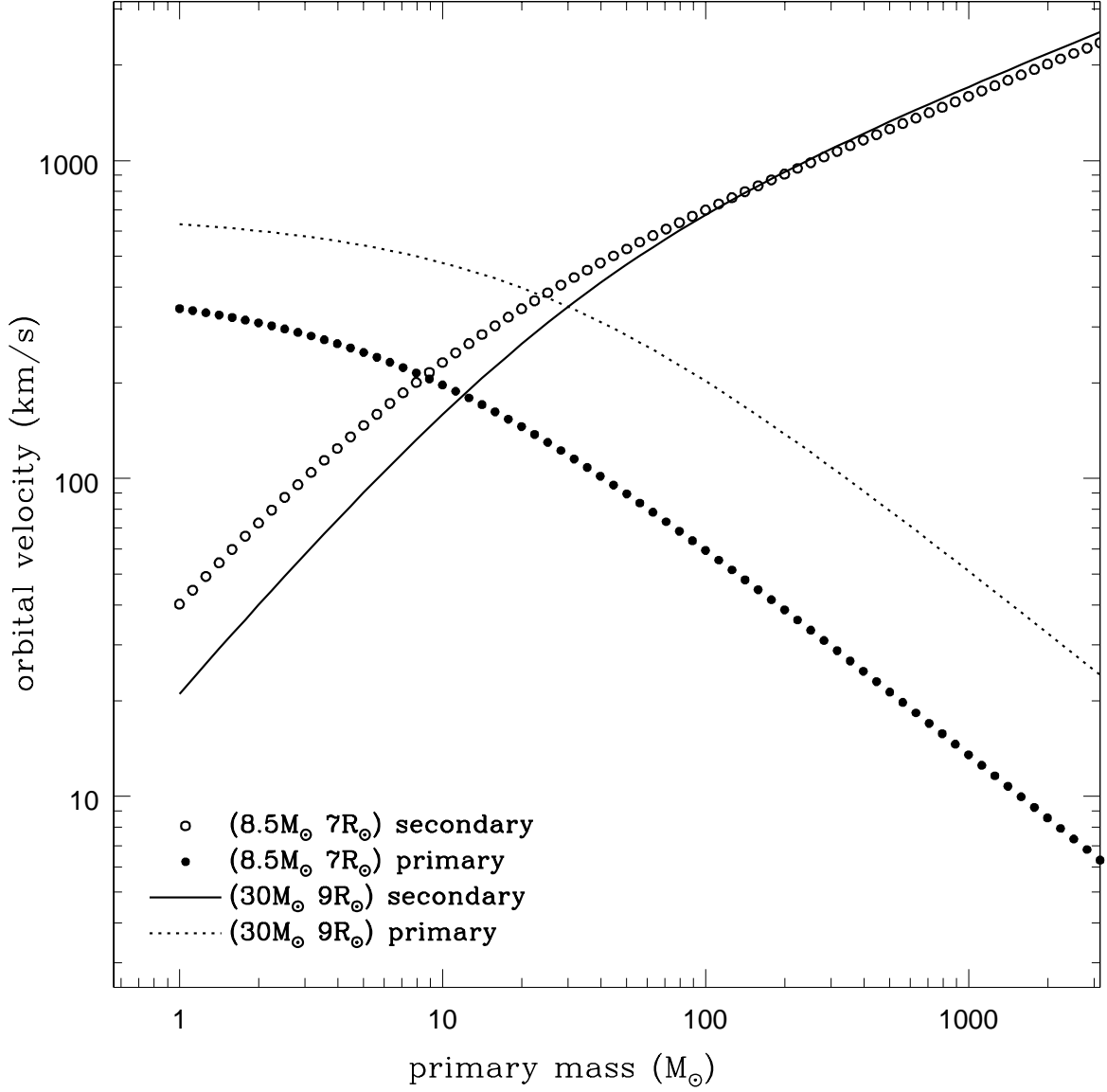}

\caption{The orbital velocities for the primary and secondary around the mass
center in NGC1313 X-2. A $Z=0.2Z_\odot$ secondary (8.5 $M_\odot$, 7 $R_\odot$)
and an O7V secondary ($30 M_\odot, 9 R_\odot$) are considered for the system.
}

\end{figure}

\begin{figure}
\plotone{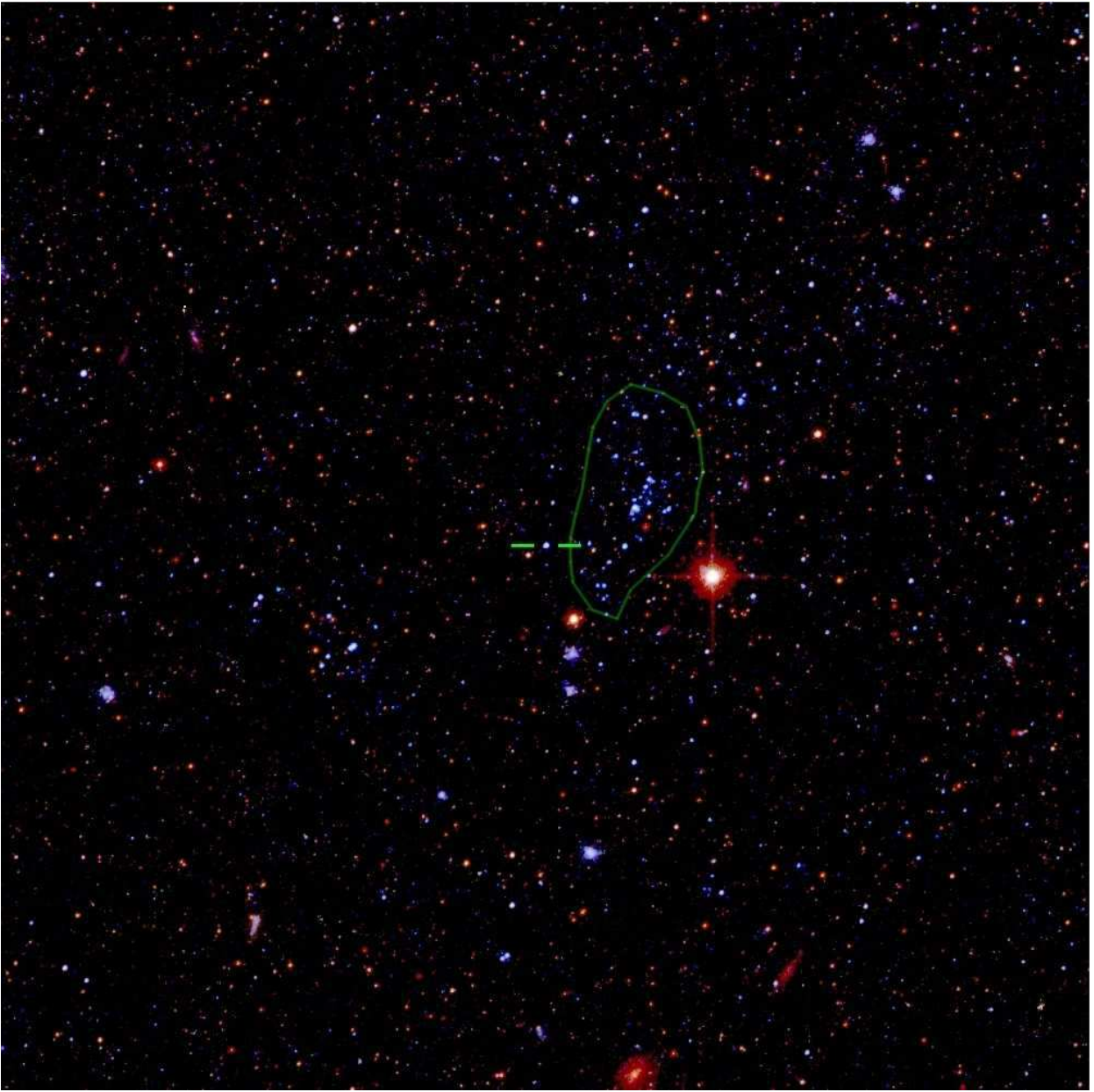}
\caption{True color image for a $50^{\prime\prime} \times 50^{\prime\prime}$
region centered at the counterpart for NGC1313 X-2. The image was composed with
F435W,F555W,F814W images from the 2003 observation. The stellar field is dominated by red stars 
with scattered blue stars. A few tens of bright blue stars are clustered
within an irregular shape of $10^{\prime\prime} \times 5^{\prime\prime}$ (i.e.,
$180pc\times 90pc$) to the west of the counterpart.  North is $7.^\circ4$
clockwise from upward vertical.  }

\end{figure}

\end{document}